\documentclass[trackchanges]{aastex701}
\usepackage{multirow}
\usepackage{booktabs}
\usepackage{rotating}
\usepackage{xcolor}
\usepackage{makecell}

\begin{document}

\title{Vertical Structure of Local Disk Galaxies revealed by DESI Imaging Data}

\author{Nan Chen}
\affiliation{Department of Astronomy, School of Physics and Astronomy, Yunnan University, Kunming, Yunnan 650500, People’s Republic of China}
\email{} 

\author[0000-0001-5258-1466]{Jianhui Lian}
\affiliation{South-Western Institute for Astronomy Research, Yunnan University, Kunming, Yunnan 650091, People’s Republic of China}
\affiliation{Yunnan Key Laboratory of Survey Science, Yunnan University, Kunming, Yunnan 650500, People’s Republic of China}
\email{jianhui.lian@ynu.edu.cn} 

\author{Yuze Zhao}
\affiliation{South-Western Institute for Astronomy Research, Yunnan University, Kunming, Yunnan 650091, People’s Republic of China}
\email{} 

\begin{abstract}
The vertical structure of galactic disks are an important probe to the disk assembly history. 
Here we investigate the vertical structure of a sample of 79 local disk galaxies within 50~Mpc using data from the DESI Legacy Imaging Surveys. Vertical luminosity profiles as a function of radius in the $g$, $r$, and $z$ bands are extracted and fitted using a single-component sech$^2$ model\ to determine scale height and its radial variation (i.e. flaring). Our measurements indicate local galactic disks are overall thin with negligible flaring. The median scale heights at 1~$R_e$ of are 0.21, 0.22, and 0.22 kpc in the $g$, $r$, and $z$ bands, respectively, while the median radial gradient of scale height are of −0.006, 0.003, and 0.001 in these bands. These values are consistent with that of the geometric thin disk of the Milky Way represented by metal-rich, low-[$\alpha$/Fe] populations, confirming the weak flaring of geometric thin disk.
A clear positive correlation of scale height with stellar mass is observed down to low mass end of 10$^7\ {\rm M_{\odot}}$.
These results provide a homogeneous benchmark for the vertical structure of nearby disk galaxies and establish the cosmological representativeness of the Milky Way thin disk.

\end{abstract}

\keywords{\uat{Disk galaxies}{391} --- \uat{Galaxy photometry}{611} --- \uat{Milky Way Galaxy}{1054}}


\section{INTRODUCTION}
The vertical structure of an astrophysical disk provides critical constraints on its structural and dynamical state and its potential interactions with the surrounding environment. For galaxies, studying the vertical structure is an effective way to understand the formation and evolution of galactic disks, as well as the underlying astrophysical processes that usually involve interactions with external galaxy companions and among internal substructures \citep{villumsen1985,quinn1993,brook2004,villalobos2008,Bournaud_2009,yi2023}.


Over the past several decades, significant progress has been made in understanding the vertical structure of the Milky Way, particularly with the discovery of the thin and thick disk components in the galactic plane. 
In a pioneering work, \citet{gilmore1983} revealed that the Milky Way disk is composed of two major components with distinct thicknesses. 
Further studies have shown that these two disk components also differ in many other properties, e.g., 
chemical abundances, stellar age, and dynamical properties \citep{fuhrmann1998,reddy2006,haywood2013,bovy2012}. More recently, a young thick disk component of particular interest with a similar vertical structure to the canonical high-$\alpha$ thick disk was identified, characterized by low-metallicity and low-$\alpha$ abundances \citep{lian2025a}. It is found both in observations and simulations that the thickness (i.e., scale height) of a disk is not constant but generally increases with radius, a phenomenon commonly referred to as “flaring” \citep{lopez-corredoira2002,momany2006,Minchev_2015,Sotillo-Ramos2023}. Thicker disk components tend to flare more strongly \citep{lian2025a}. Therefore, the vertical structure of the Galactic disk is characterized by two parameters: the scale height at a certain radius and its variation with radius. 

More recently, the advent of many massive stellar spectroscopic surveys that have delivered precise measurements of chemical abundances, ages, and distances for millions of stars has enabled detailed studies on Galactic vertical structure far beyond the solar vicinity and extending back to different epochs in the Galactic evolutionary history. This type of study is conducted by analyzing the structure of mono-abundance/age populations \citep{bovy2012,Bovy_2016,mackereth2017,Xiang_2018,Yu_2021,lian2022,Imig_2023}. These studies revealed substantial variation in the vertical structure of the disk, including the scale height at the solar radius and the strength of flaring, among different mono-abundance populations, suggesting significant structural evolution of the Milky Way over cosmic time. 

While extensive studies have been carried out for the Milky Way, explorations on the vertical structure of external galaxies remain sparse. 
Observations of edge-on galaxies have revealed that many external galaxies exhibit a similar disk structure to the Milky Way, including the presence of thin and thick disks \citep{yoachim2006,comeron2018}. High-redshift observations have further confirmed that disk thickness is already significant at early cosmic times, suggesting that the thickness of galactic disks is at least partially established in the early turbulent phase \citep{elmegreen2006,hamiltoncampos2023,lian2024}. 
Recently, \citet{Ranaivoharimina_2024} examined the flaring properties of disks in a sample of 65 local galaxies and found strongly flaring in the very outer part. 

In this work, we aim for a systematic study of the vertical structure of local disk galaxies using galaxy image data from Dark Energy Spectroscopic Instrument (DESI) Legacy Imaging Surveys. 
The paper is organized as follows. In Section \ref{sec:data}, we introduce the selection of sample galaxies and measurements of their vertical structure. In Section \ref{sec:result}, we present our main results regarding the flaring strength and scaling relations involving scale height of local galaxies. We also include the corresponding measurements in the Milky Way for an intriguing comparison. Section \ref{sec:summary} provides a summary of the study.

\section{DATA AND METHODS}
\label{sec:data}


\subsection{Sample Selection}
Our galaxy sample is selected from a catalog of 15,424 nearby galaxies within 50~Mpc \citep{2024catalog} aiming for which the $g$, $r$, and $z$ band imaging data from the Dark Energy Spectroscopic Instrument (DESI) Legacy Imaging Surveys are available. The DESI surveys provide deep, wide-field optical-to–near-infrared imaging with uniform photometric calibration and sky subtraction, making them suitable for studies of galaxy structure. We obtained images 
from DESI Data Release 10 (DR10) \citep{dey2019}.

Since the survey did not cover the full catalog, this resulted in an initial subset of 12,622 galaxies. Among these, 12,375 galaxies had complete sets of $g$, $r$, and $z$-band images required for multi-band analysis. 
{To ensure that the target galaxy is not affected by nearby bright objects, we implemented a rather strict selection criteria to exclude those with nearby saturated pixels in DESI images. 
For each galaxy, we extracted an image cutout from DESI DR10 website\footnote{https://www.legacysurvey.org/} with adjustable image size that is mildly larger than the target galaxy. We then examined the images and excluded those with saturated pixels within 2/5 image size from the galaxy center. 
This results in a sample of 5,485 galaxies.} 
To identify edge-on galaxies, we used the $r$ band images to measure the axis ratio of each galaxy. The procedure used to derive the axis ratio was as follows:

First, we applied a high detection threshold, defined as the local background level plus 20 times its root mean square (RMS), to deblend the image and isolate a compact central region of the galaxy. The pixel with the maximum flux within this region was taken as the galaxy center. Foreground stars were also identified using this high-threshold. For each detected star, we constructed a circular mask with a radius corresponding to the distance at which its surface brightness falls to the local background RMS level. The point spread function (PSF) of each star was characterized by fitting a Gaussian profile, and the resulting model was then used to determine the appropriate masking radius.
Using the pre-masked image, we recalculated the global background level and its RMS. A second, lower detection threshold—set to the level of the local background RMS—was then applied to perform more thorough deblending and to segment the full spatial extent of each galaxy. From this segmentation, we derived global structural parameters, including the axis ratio ($b/a$) and position angle (PA). Any remaining non-target sources were masked.
The final outputs for each galaxy include the central pixel coordinates, axis ratio, position angle, and a fully masked, background-subtracted image.

We select 79 edge-on disk galaxies from the parent sample to examine their vertical structure.
The criterion used to determine whether a galaxy is edge-on is based on the axis ratio of the galaxies we measured. 
We selected galaxies with an axis ratio smaller than 0.2 in $r$ band. 
Subsequently, we performed a visual inspection of the galaxy images, discarding those with excessively low inclinations (2), low signal-to-noise ratios (6), significant substructures (3), or contamination from bright foreground stars (1). Following this process, the remaining numbers of galaxies in each band are as follows: 75 galaxies in the $g$ band, 73 galaxies in the $r$ band, and 73 galaxies in the $z$ band. The galaxy samples are not completely consistent across the three bands, with 66 galaxies common to all three bands.

\begin{figure}
    \centering
    \includegraphics[width=0.5\linewidth]{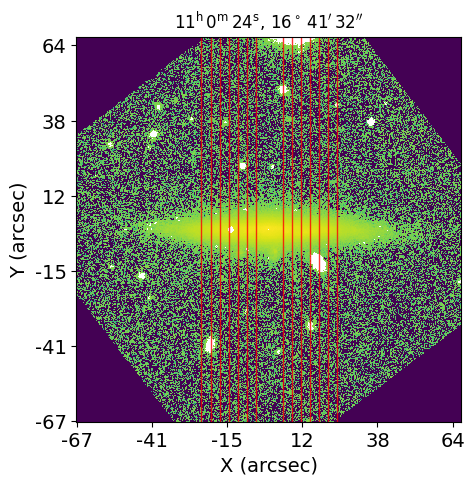}
    \caption{$r$ band image of the galaxy UGC 06083 (RA = $11^\mathrm{h} 0^\mathrm{m} 24^\mathrm{s}$, DEC = $16^\circ 41' 32''$). The red vertical lines indicate the radial range from $0.3\,R_\mathrm{e}$ to $1.5\,R_\mathrm{e}$ used to extract the vertical surface brightness profile.}
    \label{fig:1}
\end{figure}

We make an image cutout centered on the galaxy and rotate it such that the galaxy's disk is aligned with the x-axis using the position angle (PA) data derived from source detection. 
Figure \ref{fig:1} shows the processed image cutout of one of the sample galaxies, UGC 06083. The red vertical lines indicate the range from which the vertical surface brightness profile will be extracted in the next section.

\label{sub:sample section}

\label{sub:image process}

\subsection{Vertical surface brightness profiles}

To study the vertical structure across the disk, 
we extracted multiple surface brightness profiles perpendicular to the disk plane along the edge-on galaxy disk. Taking the galaxy center as the reference point, we began at a radial distance of 0.3 effective radii ($R_e$) and extracted six parallel vertical strips out to 1.5$R_e$ at intervals of 0.2$R_e$. The effective radius($R_e$) is derived from DESI images. We first extracted the radial surface brightness profile of each galaxy along its semi-major axis from its DESI image using the photutils Python package. Subsequently, we cumulatively summed these fluxes from the inner to outer regions until the total flux saturated, and identified the semi-major axis radius at which half of the total flux is enclosed. This radius is adopted as the effective (half-light) radius($R_e$) of the galaxy.
Within each strip, we merged pixels to enhance signal-to-noise ratio (SNR) while maintaining sufficient spatial resolution. For galaxies with an image scale of 1000 pixels, we merged every 8 pixels within the ranges of 0–486 pixels and 514–1000 pixels in the outskirt, and every 4 pixels within the range of 486–514 pixels close to the disk plane. For galaxies with an image scale of 512 pixels, we merged every 4 pixels within the range of 0–240 pixels, every 2 pixels within the range of 240–270 pixels, and every 4 pixels within the range of 270–512 pixels. Data points with SNR$<5$, or those disconnected from the bright central part are excluded. 

\begin{figure}
    \centering
    \includegraphics[width=1\linewidth]{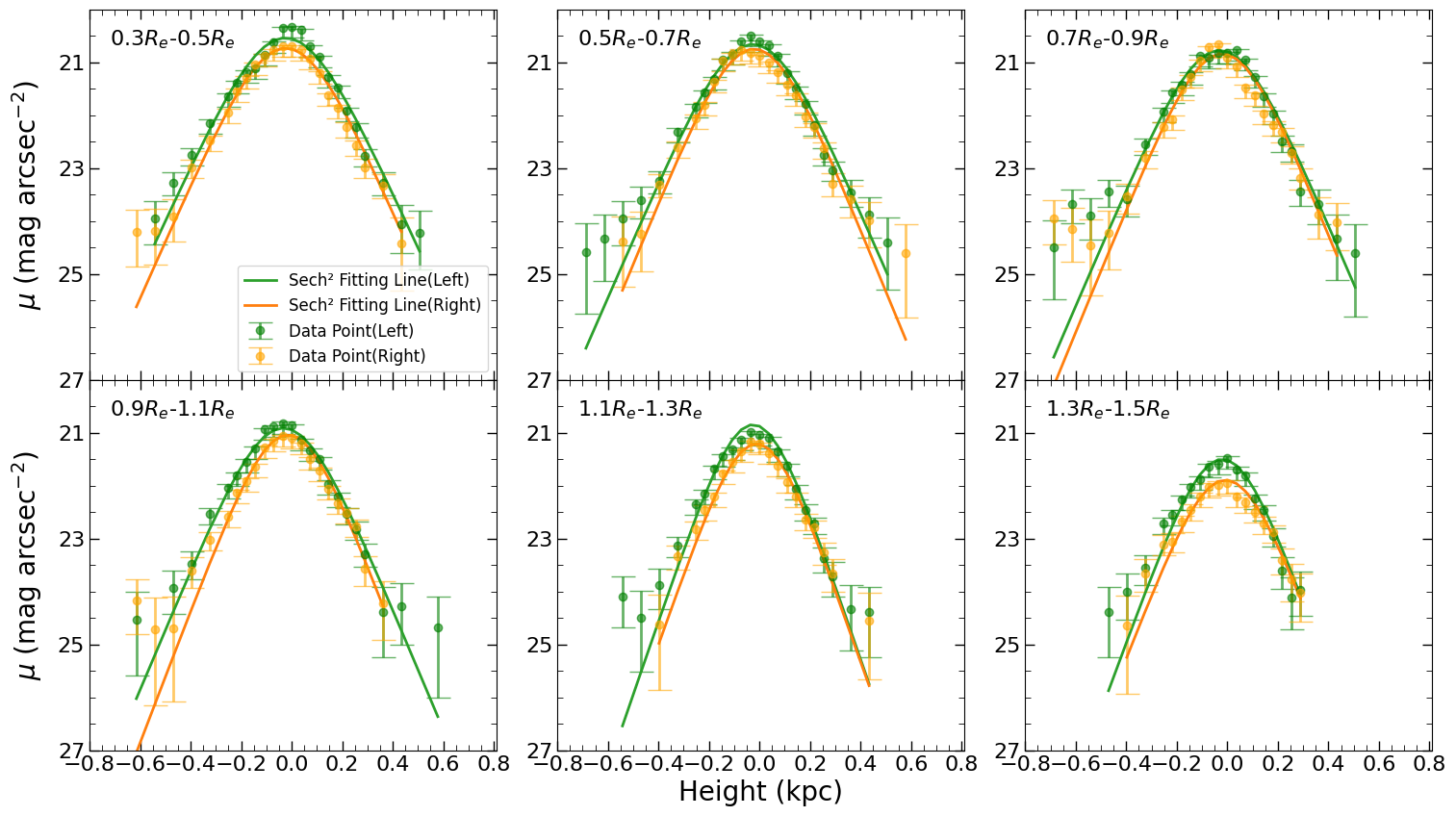}
    \caption{Vertical surface brightness profiles of the example galaxy. Each panel indicates the profile in each radial bin, ranging from 0.3$R_e$ to 1.5$R_e$ with bin width of 0.2$R_e$. 
    The green and orange dots represent the profiles at the left and right sides, respectively, while the solid lines shows the corresponding best-fitting models. Error bars indicate the uncertainties of the surface brightness measurements.}
    \label{fig:2}
\end{figure}

Figure~\ref{fig:2} shows the extracted vertical profiles at six radial bins for an example galaxy. The image depth reaches $\sim$25~${\rm mag/arcsec^2}$ and in most cases only geometrically thin disk can be identified. This is not inconsistent with the result of the prevalence of thick disk in local star-forming galaxies \citep{comeron2018} as it normally requires deep imaging down to 27~${\rm mag/arcsec^2}$. In light of this, we fit the extracted vertical profile at each radii with a single-component sech$^2$ model \citep{kruit1981} as follows: 

\begin{equation}
    \label{eq:公式1}
    \mu(x)=\frac{\mu_0}{\cosh\left(\frac{|x - x_0|}{2h_z}\right)^2} 
    \end{equation}

Where $\mu_0$ is the central surface brightness, $h_z$ denotes the scale height, which is comparable to the exponential scale height, and $x_0$ is the coordinate of the peak of the surface brightness profile. Prior to fitting the data, the density model is convolved with a simple PSF model represented by a Gaussian kernel with FWHM of 1.5\arcsec which is representative of the seeing conditions when taking the images. We use the curve-fit function from SciPy package to perform non-linear least-squares fitting. Each data point is weighted by its inverse variance of the surface brightness measurement.

We performed the above operations on both sides of the galaxy disk, yielding 12 fitted curves, as shown in Figure \ref{fig:2}. These 12 fitted curves correspond to the vertical surface brightness distributions at 0.4 $R_e$, 0.6 $R_e$, 0.8 $R_e$, 1.0 $R_e$, 1.2 $R_e$, and 1.4 $R_e$ on each side. 
We averaged the two sets of values from left and right side to obtain the $h_z$ and $\mu_0$ for each galaxy at each radii. The errors for $h_z$ and $\mu_0$ are derived from Monte Carlo simulation with 100 trials of resampling and fitting. 

Due to the varying inclination angles of edge-on galaxies, the measurement of their scale heights tends to be overestimated \citep{hamiltoncampos2023}. To address this, we used IMFIT \citep{Erwin_2015} to construct a set of edge-on model galaxies with different inclination angles (ranging from 0 to 12 degrees with a step size of 0.1 degrees). We then re-measured the scale heights of these model galaxies and compared them with the original values we set, calculating the percentage by which the scale height was overestimated. The median percentage of overestimation from this set of model galaxies was found to be 8.01\%. We applied this correction factor to all our scale height results to account for the overestimation introduced by the inclination angle.

\label{sub:vertical}

\section{RESULT AND ANALYSIS}
\label{sec:result}


\subsection{Disk thickness and flaring}

\begin{figure}
    \centering
    \includegraphics[width=1\linewidth]{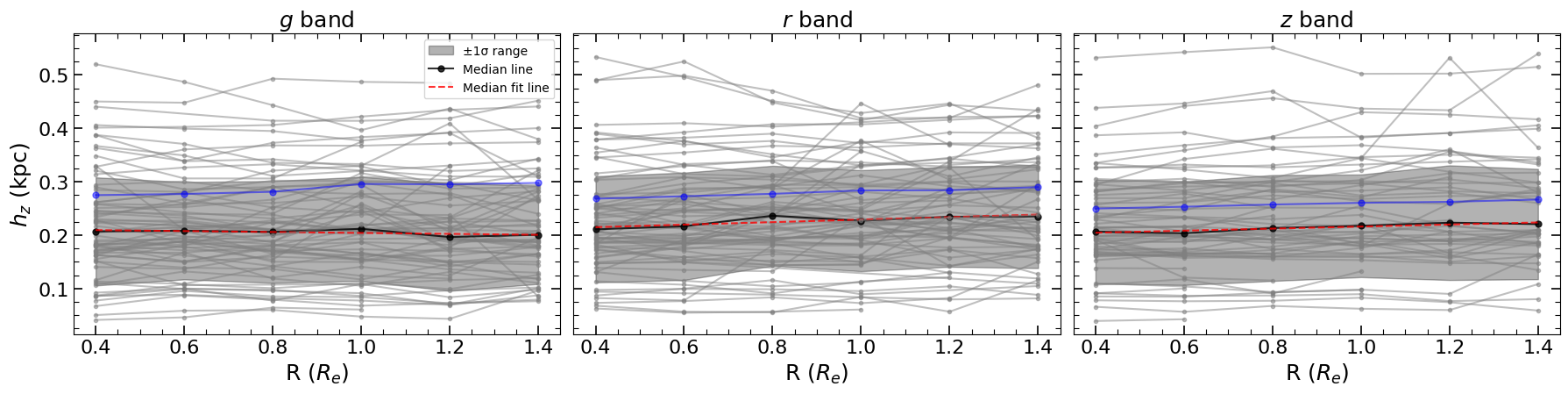}
    \caption{The scale height ($h_z$) as a function of distance from the galaxy's center in unit of $R_e$ for three bands: $g$, $r$, and $z$. Gray circles and lines represents scale heights of individual galaxies, while black circles indicate the median value at each radii, with the red dashed line showing the the best-fitted model to the median values. The gray shaded area indicates the $\pm 1\sigma$ range. The blue line highlights an example galaxy (NGC 4445) exhibiting non-zero thin-disk flaring.}
    \label{fig:3}
\end{figure}

Figure \ref{fig:3} shows the obtained scale heights $h_z$ as a function of radius for all galaxies across the three bands. The scatter of the distribution is depicted by the shaded areas, and the black dots represent the median values for each radial bin. 
Notably, in all three bands, the scale heights remain roughly constant across the radial range, without significant radial variation. This suggests that the overall vertical structure of local galaxies' disks i.e., exhibits negligible flaring within 1.5~$R_e$.

\begin{figure}
    \centering
    \includegraphics[width=1\linewidth]{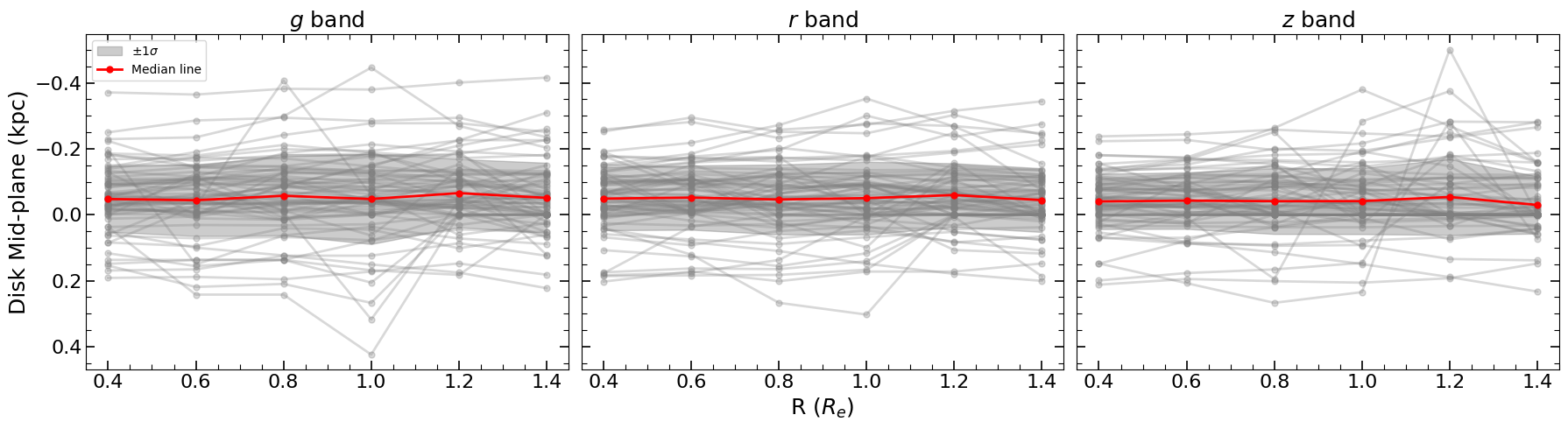}
    \caption{Variation of the disk mid-plane height as a function of $R_e$ in the $g$, $r$, and $z$ bands. Gray thin lines represent individual galaxies, with gray points showing the measured mid-plane positions at each radial bin. The red solid line with circular markers indicates the median trend at each radius. The gray shaded area indicates the $\pm 1\sigma$ range. In all three bands, the median mid-plane remains close to zero, showing no strong systematic warp signal. Only mild deviations appear at larger radii, and the overall behavior is consistent across the different photometric bands.}
    \label{fig:4}
\end{figure}

\begin{figure}
    \centering
    \includegraphics[width=1\linewidth]{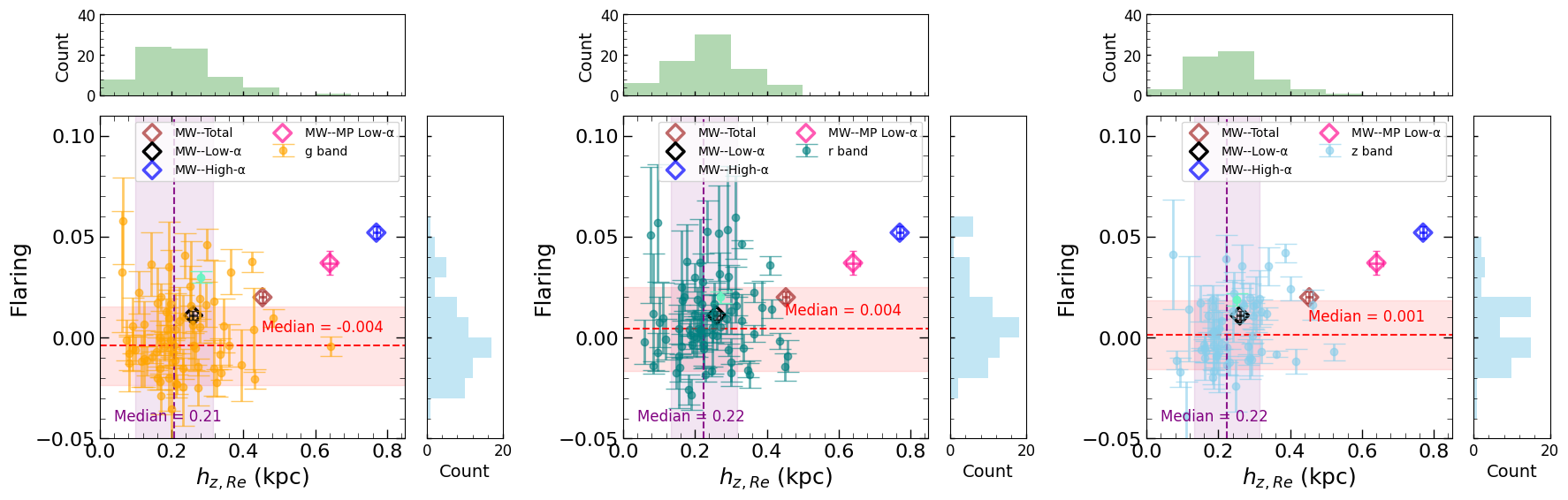}
    \caption{Galaxy distribution in flaring strength and scale height at $R_{\rm e}$ ($h_{z, R_{\rm e}}$). Three panels shows the distribution at three bands: $g$ (left), $r$ (middle), and $z$ (right). Top and right histograms show the distribution in flaring strength and $h_{z, R_{\rm e}}$, respectively. The median values of these two properties are marked as horizontal and vertical red dashed lines. Stars with different colors represent the measurements of the Milky Way's disk in different components: total (brown), high- and low-$\alpha$ (blue and black), and metal-poor low-$\alpha$ (pink). The fluorescent green points highlight a representative galaxy (NGC 4445) exhibiting relatively strong thin-disk flaring.}
    \label{fig:5}
\end{figure}

\begin{table}
\centering
\caption{Galaxy Flaring and Scale Height at 1$R_{\rm e}$}
\label{table}
\begin{tabular}{lcc|ccc|ccc}
\toprule
\multirow{2}{*}{Name} &
\multirow{2}{*}{RA} & 
\multirow{2}{*}{DEC} &
\multicolumn{3}{c|}{Flaring ($k$)} &
\multicolumn{3}{c}{h$_z$ at 1$R_{\rm e}$} \\
\cmidrule(lr){4-6} \cmidrule(lr){7-9}
 & (deg) & (deg) & $g$ & $r$ & $z$ & $g$ & $r$ & $z$ \\
(1) & (2) & (3) & (4) & (5) & (6) & (7) & (8) & (9) \\
\midrule
PGC090975 & 150.29229 & 0.90851 & 0.03 ± 0.06 & 0.05 ± 0.06 & 0.03 ± 0.04 & 0.06 ± 0.01 & 0.08 ± 0.01 & 0.07 ± 0.00 \\
PGC090973 & 150.47939 & 63.89265 & 0.03 ± 0.03 & 0.02 ± 0.05 & 0.02 ± 0.09 & 0.11 ± 0.01 & 0.14 ± 0.02 & 0.11 ± 0.01 \\
\bottomrule
\end{tabular}
\tablecomments{The full table is available in the online supplementary material. Column (1): galaxy name.
Column (2): right ascension at equinox J2000.
Column (3): declination at equinox J2000.
Column (4–6): flaring measured in the $g$, $r$ and $z$ bands, derived from the radial variation of the vertical scale height.
Column (7–9): scale height $h_z$ at $1R_e$ in the $g$, $r$ and $z$ bands, obtained from the best-fitting vertical luminosity profiles.}
\end{table}

To further investigate the disk flaring of individual galaxies, we fit a linear model to the scale heights as a function of radius for each galaxy, with the slope representing the strength of flaring. The fitting results in the $g$, $r$, and $z$ bands are shown in Figure \ref{fig:5}, where the vertical axis represents the unitless flaring strength and the horizontal axis represents the scale height at $1~R_{\rm e}$ (${\rm h_{z, R_{\rm e}})}$. The histograms shown in the upper and right sections of each plot highlight the distribution of flaring strength and scale height at $1~R_{\rm e}$. In the $z$ band, there is an outlier located in the lower-left corner of the plot. This point corresponds to a galaxy whose right half of the image is missing due to contamination from a foreground star, resulting in a relatively large measurement error. It can be seen that the majority of local galaxies have relatively small flaring values, with most showing flaring strengths close to zero across the three bands. The median flaring values for each band are -0.004, 0.004, and 0.001 in the $g$, $r$, and $z$ bands, respectively. The negative flaring values shown in Figure \ref{fig:5} are primarily caused by measurement uncertainties. Among the three bands, the $r$ band has the highest signal-to-noise ratio and data quality, resulting in a smaller dispersion. The detailed measurements for individual galaxies, including RA, DEC, flaring strengths and scale heights ($h_z$) with associated errors in all three bands, are listed in Table\ref{table}.

\citet{Ranaivoharimina_2024} derived the scale heights as a function of radius out to 0.8~$R_{25}$($R_{25}$ is defined as the galactocentric radius corresponding to the isophote at a surface brightness level of $\mu$ = 25 mag $arcsec^{-2}$ \citep{trujillo2020}) for a sample of 65 nearby galaxies. They found notable flaring of the outer disk beyond 0.6~$R_{25}$, while the disks within 0.4~$R_{25}$ generally exhibit weak flaring. This is qualitatively consistent with our result as the measurements presented here are confined to the galactic disks within 1.5~$R_{e}$, which roughly corresponds to  0.45~$R_{25}$.  

For reference, we include the measurements of flaring for different components of the Milky Way \citep{lian2022}. The Milky Way's stars exhibits two sequences in [$\alpha$/Fe]-[Fe/H] diagram with distinctive $\alpha$ abundances. These two sequences can be roughly separated at [$\alpha$/Fe] $\approx$ 0.2 dex, with stars below (above) this threshold classified as low-$\alpha$ (high-$\alpha$) populations \citep{Adibekyan2011}. We include the flaring properties of the integrated low-$\alpha$ and high-$\alpha$, and metal-poor low-$\alpha$ populations of the Milky Way for comparison \citep{lian2025a}. The last one represents a newly identified young thick disk substructure of the Milky Way.
The Milky Way's disk, when treating as a single component, is mildly thicker with stronger flaring comparing to local disk galaxies. Among the various components, the low-$\alpha$ disk exhibits comparable low thickness and weak flaring, whereas the old and young thick disk components show much greater thickness and stronger flaring than local galactic disks. 
The consistency between measurements of local galactic thin disk and the low-$\alpha$ disk in the Milky Way suggests that the vertical structure of the thin disk in the Milky Way is typical among the local galaxy population. Additionally, it reinforces the trend of weaker flaring in thinner disk reported in \citet{lian2025a}, suggesting the flaring and thickness of galactic disks share a common origin.

In addition, we examine the possible presence of disk warping. In our fitting procedure, $x_0$ was treated as a free parameter, and its absolute value represents the magnitude of the disk mid-plane height. As shown in Figure \ref{fig:4}, within the radial range explored in this study (0.4–1.4 $R_e$), the mid-plane height does not exhibit significant systematic variations, indicating no detectable warp signature. For the Milky Way, the disk warp becomes prominent at galactocentric distances of 11–14 kpc\citep{chen2025}. In contrast, the maximum radius considered in our analysis is 1.4 $R_e$, corresponding to 8.05 kpc (assuming $R_e$ = 5.75 kpc for the Milky Way), which lies well inside the typical warp region. Observational studies of nearby galaxies further show that disk warps generally occur at radii of 2–3 disk scale lengths, typically beyond ∼2 $R_e$ \citep{Reshetnikov2016,Saha2009}. Therefore, our investigation of disk flaring is not expected to be affected by disk warping.

\subsection{Mass Dependence}
\begin{figure}
    \centering
    \includegraphics[width=1\linewidth]{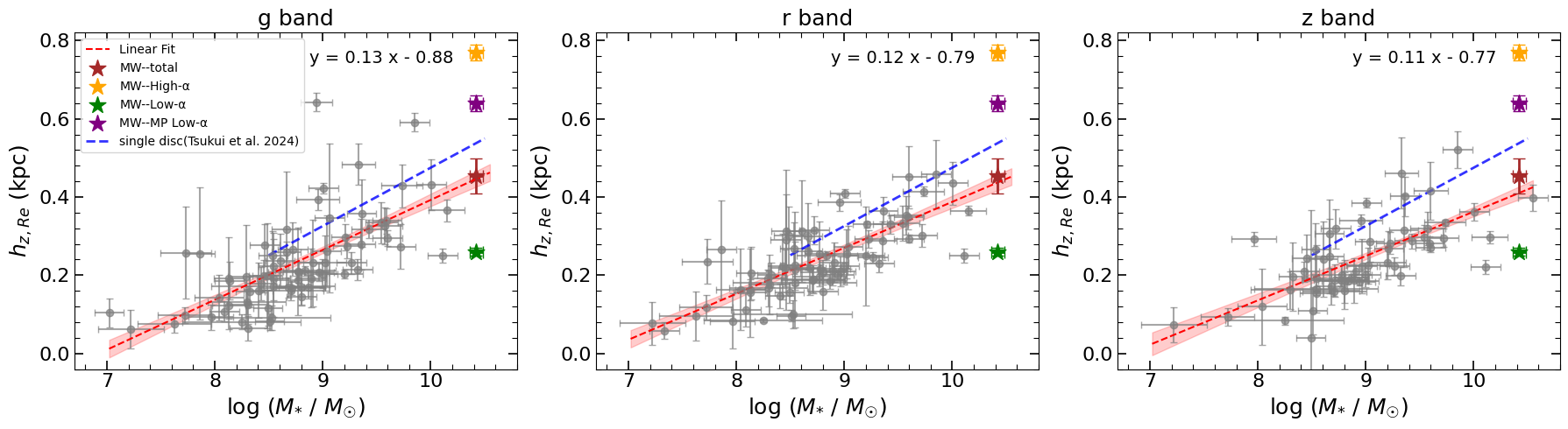}
    \caption{Mass dependence of disk thickness of local galaxies. Grey dots show the measurements of individual galaxies. 
    Red dashed lines are the best linear fit to the data points, with the shaded region for 1$\sigma$ uncertainty derived from Monte Carlo (MC) simulation. Blue dashed line shows the `single disk' fitting result for distant galaxies from \cite{tsukui2025emergencegalacticthickdiscs}, which is repeated in the three panels for reference. Stars represent the measurements of the Milky Way's disk in different components. 
    }
    \label{fig:6}
\end{figure}

Here we explore the mass dependence of the disk thickness based on our fitting results.
Figure~\ref{fig:5} shows the scale height at 1~$R_{\rm e}$ in three bands as a function of stellar mass. A strong correlation between stellar mass and disk thickness is observed, with more massive galaxies generally being thicker. This correlation holds in all three bands and down to low mass end of $10^7\ {\rm M_{\odot}}$. We fit a linear model to the mass-scale height relation with the best-fitted results presented in Figure~\ref{fig:6}. The mass dependence is similar in all three bands, with $g$ band being slightly steeper. 
A similar mass dependence has been reported in distant galaxies \citep{elmegreen2006,tsukui2025emergencegalacticthickdiscs}. 
Our results suggest it persists to very low mass galaxies. The mass-scale height relation for galaxies fitted with `single-disk' model in \citet{tsukui2025emergencegalacticthickdiscs} is included for comparison. Our best-fitted model is overall consistent with the relation derived for more distant galaxies, albeit with slightly lower scale height. Despite differences in sample selection and measurement methodology, the slopes we derive in all three bands closely match their results, reinforcing the view that galaxy mass plays a in determining disk thickness. 

Similar to flaring, we also include the scale height measurements of the Milky Way’s disk for comparison. For simplicity, for different disk components we use the same stellar mass of $2.603\times10^{10}\ {M_{\odot}}$ from \citet{lian2025b} that accounts for the broken density profile of the Galactic disk. Unlike flaring, the scale height of the Milky Way's disk as a whole is more consistent with our galaxy sample, while the low-$\alpha$ disk is mildly thinner and the high-$\alpha$ and metal-poor low-$\alpha$ disks are substantially thicker. We note that this comparison is unfair for individual disk components, as their scale heights should correlate more directly with their own stellar masses. Future wide-area deep imaging surveys (e.g., Euclid, CSST, and Roman) will allow decomposition of local disk galaxies into different components and more direct component-by-component comparison with the Milky Way.

\section{Summary}
In this work, we investigate the vertical structure of local disk galaxies using data from the DESI Imaging Legacy Survey.
We identify 75 edge-on galaxies in the $g$ band, 73 in the $r$ band, and 73 in the $z$ band with axis ratios smaller than 0.2. We extracted vertical luminosity profiles in six radial bins from 0.3 to 1.5 $R_{\rm e}$ of these galaxies and fit them using a single-component sech$^2$ model given no clear detection of secondary thick component. 
A correction for the overestimation of scale height caused by non-perfect inclination has been taken into account. 

We then investigate the radial variation and mass dependence of the disk thickness in these galaxies and compare them with the Milky Way. Our main conclusions include:
\begin{itemize}
    \item The scale heights of these local galaxy disks are generally thin with median values of 0.21, 0.22, and 0.22kpc in $g, r$, and $z$ band, respectively. These values are similar to those of the low-$\alpha$ thin disk of the Milky Way.
    \item The flaring of local galaxy disks is generally weak. The median flaring (defined as radial gradient of scale height) in the $g$, $r$, and $z$ bands are -0.004, 0.004, and 0.001, respectively, comparable to that of the overall low-$\alpha$ thin disk of the Milky Way, but significantly lower than the high-$\alpha$ and metal-poor low-$\alpha$ thick disks.    
    \item A strong correlation between the disk thickness and galaxy stellar mass is found. More massive galaxies are systematically thicker. This relationship is valid over three orders of magnitude in stellar mass from 10$^7\ {\rm to\ 10^{10} M_{\odot}}$. 
\end{itemize}



Our results reveal typical local galaxy disks with short scale height and weak flaring, which are consistent with the geometric thin disk of the Milky Way represented by metal-rich, low-$\alpha$ abundance. This suggests that the Milky Way's thin disk is typical in the local Universe and likely shares a similar formation and evolutionary history with local galaxies.

In the future, we anticipate that the ongoing and upcoming wide-area deep imaging surveys conducted by ground- (e.g., LSST) and space-based telescopes (Euclid, CSST, and Roman) will open the opportunity to ubiquitously detect and study the faint thick disk of local galaxies and establish a more direct connection with the knowledge gained from our home galaxy, the Milky Way.  
\label{sec:summary}

\section*{Acknowledgements}
This work is supported by National Key R\&D Program of China (No. 2024YFA1611600), National Natural Science Foundation of China (No.12473021), Yunnan Province Science and Technology Department (No. 202105AE160021 and 202005AB160002), and Key Laboratory of Survey Science of Yunnan Province (No. 202449CE340002). 

This work makes use of data obtained with the Dark Energy Spectroscopic Instrument (DESI).DESI construction and operations is managed by the Lawrence Berkeley National Laboratory. This research is supported by the U.S. Department of Energy, Office of Science, Office of High-Energy Physics, under Contract No. DE–AC02–05CH11231, and by the National Energy Research Scientific Computing Center, a DOE Office of Science User Facility under the same contract. Additional support for DESI is provided by the U.S. National Science Foundation, Division of Astronomical Sciences under Contract No. AST-0950945 to the NSF’s National Optical-Infrared Astronomy Research Laboratory; the Science and Technology Facilities Council of the United Kingdom; the Gordon and Betty Moore Foundation; the Heising-Simons Foundation; the French Alternative Energies and Atomic Energy Commission (CEA); the National Council of Science and Technology of Mexico (CONACYT); the Ministry of Science and Innovation of Spain, and by the DESI Member Institutions. The DESI collaboration is honored to be permitted to conduct astronomical research on Iolkam Du’ag (Kitt Peak), a mountain with particular significance to the Tohono O’odham Nation.

\bibliography{nancy_check}{}
\bibliographystyle{aasjournalv7}

\end{document}